\documentclass[a4paper,showpacs,twocolumn,english,aps,prl]{revtex4}
\usepackage[T1]{fontenc}
\usepackage[latin1]{inputenc}
\usepackage{float}
\usepackage{graphicx,pictex}

\usepackage{amsfonts,multirow,amsmath,xspace} %

\usepackage[%
  colorlinks=false %
, pdfpagemode=None %
, hypertex %
]{hyperref} %

\makeatletter

\usepackage{babel}
\makeatother
\begin{document}

\preprint{This line only printed with preprint option}

\title{Thermodynamics of localized magnetic moments in a Dirac conductor}

\author{V.~Cheianov, M.~Szyniszewski, E.~Burovski, Yu.~Sherkunov, and V. Fal'ko}

\affiliation{Physics Department, Lancaster University, Lancaster, LA1 4YB,
UK}

\begin{abstract}
We show that the magnetic susceptibility of a dilute ensemble of magnetic impurities in a conductor with a 
relativistic electronic spectrum is non-analytic in the inverse tempertature at $T^{-1}\to 0$.
We derive a general theory of this effect and construct the high-temperature expansion for the disorder 
averaged susceptibility to any order, convergent at all tempertaures down to a possible ordering transition.
When applied to Ising impurities on a surface of a topological insulator, the proposed general theory agrees 
with Monte Carlo simulations, and it allows us to find the critical temperature
of the ferromagnetic phase transition.
\end{abstract}
\maketitle

Collective phenomena in random ensembles of magnetic impurities embedded 
in  metallic conductors are caused by the long-distance exchange interaction mediated 
by the mobile carriers, also known as the RKKY exchange \cite{RKKY}. 
The RKKY interaction has a well studied universal structure for all metallic systems \cite{RKKYClass}, 
with the exception of the recently discovered class of two dimensional (2D)
materials in which the low-energy electron excitations resemble massless Dirac particles: 
graphene \cite{GrapheneDiscovery,Castro}, chiral metals formed at the surface of 
topological insulators \cite{ChiralMetal,ChiralMetal1}, and  silicene \cite{silicene}.
Recent experiments \cite{Chen} 
and \cite{Wray} have reported the formation of a band gap in a chiral metal contaminated 
by magnetic impurities, pointing towards magnetic ordering at the surface of the 
topological insulator; also theoretical modelling suggested ordering transition in some of 
such systems \cite{Structural,Abanin}.

\begin{figure}
\includegraphics[width=0.45\textwidth]{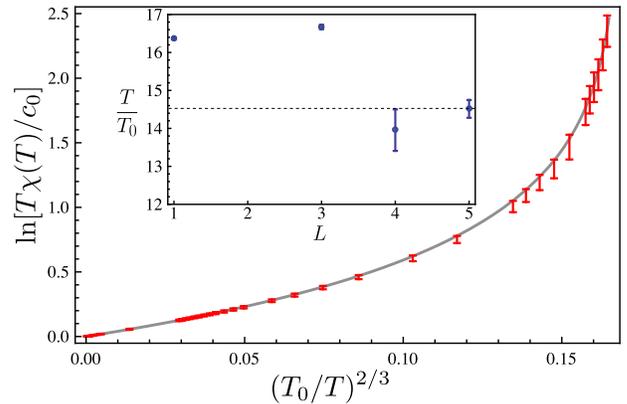}
\caption{Magnetic susceptibility of the Dirac-RKKY Ising model.
The solid line shows the L=5   D-log Pad\'e approximant calculated from the first twelve terms in the expansion \eqref{answer}
(see Table~\ref{I} for the numerical values of the coefficients $C_n$). The points 
with error bars are the results of the Monte Carlo simulation. The error bars show the statistical 
uncertainty arising from both MC fluctuations and quenched disorder. {\it Inset:} The approximate critical 
temperature derived from the position of the singularity of the D-log Pad\'e approximant. The error bars show 
the uncertainty due to the finite precision of the expansion coefficients in Table~\ref{I}. The thin dotted line shows 
the best aproximation for the critical temperature.
}
\label{Fig1}
\end{figure}

There are two peculiarities of the RKKY exchange in conductors with the Dirac-like 
electron spectrum which make it qualitatively different from usual metals: 
(i) the exchange interaction as a function of distance between 
two impurities shows the  unusual $1/r^3$ decay law, (ii)  the Friedel oscillations are 
either absent or comensurate with the lattice \cite{Graphene-Friedel}. In the following, 
we shall call such interaction a Dirac-RKKY exchange.
Other details of the RKKY exchange such as its anisotropy or 
whether it is ferromagnetic, antiferromagnetic or depolarizing may depend 
on the material and the symmetry of the impurity position in the lattice.  
In this Letter we propose a general quantitative theory of the thermodynamics 
of an ensemble of randomly positioned magnetic impurities interacting through the 
Dirac-RKKY exchange in a paramagnetic phase, that is at 
temperatures above the magnetic ordering temperature $T_c.$  
We show that, due to a peculiar decay law of the Dirac-RKKY exchange, 
the magnetic susceptibility shows a strong deviation from the 
Curie-Weiss law seen as non-analiticity in its high-temperature expansion,
\begin{equation}
\chi = \frac{c_0 }{ T} \sum_{n=1}^{\infty} C_n \left(\frac{T_0}{T} \right)^{\frac{2}{3}(n-1) } ,
\label{answer}
\end{equation}
where $c_0$ is the Curie constant, $s$ is the impurity spin quantum number, 
$T_0\propto \rho^{3/2} $ is a temperature scale dependent on the density of 
impurities, $\rho,$ and  
 $C_n$ are numerical coefficients expressed  in terms of finite-dimensional integrals of elementary
functions, see Eqs. \eqref{Cdef}, \eqref{Qdef}.  The expansion \eqref{answer} also encodes 
detailed information about the critical point of a magnetic transition: the value of $T_c$ and the susceptibility 
critical exponent $\gamma$ can be extracted from the values of several
coefficients $C_n$ with the help of Pad\'e approximation \cite{Baker1975}.
One example of a successful 
application of the proposed theory is illustrated in Fig.~\ref{Fig1}, where the susceptibility obtained with 
the help of Eq.~\eqref{answer} is compared with the Monte Carlo data for  the archetypal Dirac-RKKY 
system of  randomly positioned Ising spins with ferromagnetic $1/r^3$ exchange. 
Numerical values of the coefficient $C_2$ for some other Dirac-RKKY models are presented in Table~\ref{II}.

A pair of localized spins 
separated by a distance exceeding a few lattice constants experiences the
RKKY interaction mediated by electrons near the Fermi surface, which in Dirac conductors
consists of a discrete set of points in the reciprocal space. 
In graphene there are two Fermi points related by time reversal symmetry\cite{Castro}. 
In topological insulators where the ultrarelativistic electrons reside at the surface, 
there may be one Fermi point per face \cite{ChiralMetal}.  
The dispersion relation of the electronic excitations  near each Fermi point 
is $\epsilon(p) =\pm \hbar v p,$ where $p$ is the Bloch wave number of 
an electron relative to the Fermi point. Due to the discrete 
geometry of the Fermi sufrace and the linear dispersion of the excitations, the 
Dirac-RKKY interaction does not 
exhibit Friedel oscillations found in usual metals and decays as 
$1/r^3$ as a function of distance 
$r$ between spins \cite{Structural,Graphene-RKKY}. 
This decay law is valid in a broad range of lengths 
$r_E<r<r_T,$ where $r_{E}= \hbar v/E$ is the length scale associated 
with the binding energy $ E$  ($\sim 1$ eV) of adatom, and 
 $r_T = \hbar v/kT$ is the thermal wevelength of an electron. 
The most general form of the Dirac-RKKY Hamiltonian of a system of 
$N$ impurities in a magnetic field is 
\begin{align}
H_{12 \dots N}=
J 
\sum_{i<j}^N \frac{G_{ij}}{r_{ij}^3}- h S_z\ , 
\label{DHam} \\
\notag
G_{ij}= G(\mathbf s_i, \mathbf s_j, \mathbf n_{ij}, \xi_i, \xi_j).
\end{align}
Here the sum is taken over all pairs of impurities randomly distributed with 
the average density $\rho$ in the conductor plane, and 
$r_{ij} = \vert \mathbf r_i - \mathbf r_j\vert$ is the distance between a pair 
of adatoms with quenched positions $\mathbf r_i$ and $\mathbf r_j.$  In 
Eq.~\eqref{DHam} we assume without 
loss of generality that the magnetic 
field $h$ is coupled to the $z$-projection of the total spin $S_z=s_1^z+\dots s_N^z,$
and the spin of each impurity is assumed to be in the $2s+1$ dimensional representation 
of SU(2).  The parameter $J$ is specific for the given host material and the type of 
impurity.  Together with the impurity density $\rho$ it defines the energy scale
\begin{equation}
 T_0=\frac{s(s+1) J}{3} \rho^{3/2}.
\end{equation}
Further material-dependent details of the RKKY interaction are encoded in the dimensionless 
pairwise interaction function $G_{ij},$ which depends on two impurity 
spins $\mathbf s_{i}, \mathbf s_{j}, $
a  unit vector $\mathbf n_{ij}= \mathbf r_{ij}/r_{ij}$  
and two discrete quenched random variables $\xi_i, \xi_j$ defining the 
position of each adatom inside the lattice unit cell \cite{Structural,footnote1}
(see Table~\ref{II} for examples).
\begin{table*}
\begin{center}
 \begin{tabular}{|l | l | l | c | c |}
 \hline 
Dirac-RKKY ensemble
&
\footnotesize
$G_{ij}$
&
Observable
&
$C_2$
&
$\Theta/T_0$
\\  \hline\hline
\ Spin 1/2 Ising impurities. 
&
$ -s_{i}^z  s_{j}^z $
& 
$S_z$  
&
$9.02$
&
$27.1$ 
\\  \hline
\multirow{2}{*}{
\begin{minipage}[0pt]{6 cm}{\begin{flushleft}
Spin 1/2 impurities isotropically coupled to 
the electrons in a chiral metal.
\end{flushleft}
}
\end{minipage}
}
&\multirow{2}{*}{
$ -s_{i}^z  s_{j}^z - 
(\mathbf n \cdot \mathbf s_i)(\mathbf n \cdot \mathbf s_j)+\frac{1}{2} (\mathbf n \times \mathbf s_i)_z
(\mathbf n \times \mathbf s_j)_z $
}
& 
$S_z$  
&
$9.59$
&
$29.7$ \\
\cline{3-5}
&
&
$S_x$
&
$1.18$
&
$1.28$
\\ \hline

\multirow{2}{*}{
\begin{minipage}[0pt]{6 cm}{\begin{flushleft}
Spin 1/2 impurities with X-Y coupling to 
the electrons in a chiral metal. 
\end{flushleft}
}
\end{minipage}
}
&
\multirow{2}{*}{
$- 
(\mathbf n \cdot \mathbf s_i)(\mathbf n \cdot \mathbf s_j)+\frac{1}{2} (\mathbf n \times \mathbf s_i)_z 
(\mathbf n \times \mathbf s_j)_z $
}
&
$S_z$
&
$3.17$
&
$5.63$
\\ 
\cline{3-5}
& 
& 
$S_x$
& 
$1.17$
&
$2.16$
\\ \hline
\multirow{2}{*}{
\begin{minipage}[0pt]{6 cm}{\begin{flushleft}
Spin 1/2 impurities 
in graphene located at centres of hexagons.
\end{flushleft}
}
\end{minipage}
}
&
\multirow{2}{*}{
$ -(\mathbf s_i\cdot \mathbf s_j) \cos\frac{2\pi}{3} (\xi_i-\xi_j), \quad 
\xi\in \{ 0, 1, 2 \}
$
}
&
$S_z$
&
$-2.20$
&
$-3.27$
\\
\cline{3-5}
& 
&
$\Phi$
&
$3.90$
&
$7.69$
\\ \hline
\end{tabular}
\end{center}
\caption{Examples of Dirac-RKKY Hamiltonians and the corresponding values of the coefficients $C_1,$ $C_2$ and the generalized 
Curie temperature $\Theta.$ The two dimensional host material is assumed to be in the x-y plane. The susceptibility is calculated 
for the observable given in the third column. The observable $\Phi$ in the last row defines a staggered order  
in which impurity spins in graphene are correlated with the $\sqrt 3\times \sqrt 3$ superlattice induced by 
the commensurate Friedel oscillations \cite{Structural, Sherkunov}.
\label{II}}  
\end{table*}

{\it The non-analiticity of the high-temperature expansion}, 
Eq.~\eqref{answer}, results from the interplay between the peculiar $1/r^3$ Dirac-RKKY
interaction and the randomness of the distribution of impurities in the system. 
The susceptibility per spin of the system can be expressed in 
terms of the spin-spin correlation function, as 
\begin{equation}
\chi(T)=\frac{c_0}{NT} \sum_{i,j=1}^N \langle s^z_i s^z_j\rangle ,
\label{chidef}
\end{equation}
where the average is taken over both the thermal configurations of impurity spins with the 
Boltzmann weight defined by the Hamiltonian \eqref{DHam}, and the quenched random variables. 
The $1/r^3$ dependence of the Dirac-RKKY interaction in \eqref{DHam} makes it impossible 
to use the standard $1/T$ expansion to analyse 
the susceptibility \eqref{chidef}. Indeed, for any temperature 
$T_c<T<T_E$ the ensemble contains a finite fraction of pairs, in which 
the spins are close enough to each other to be strongly correlated.
Consider, for example, the classical ferromagnetic Ising model with 
$G_{ij}=-s_i s_j,$ where $s_i=\pm s.$ 
Due to the presence of correlated pairs the high-temperature asymptotics 
of the susceptibility splits into two contributions.
Those spins that belong to pairs smaller than the correlation radius 
$R_0= (J/T)^{1/3}=\rho^{-1/2} (T_0/T)^{1/3} $ are strongly bound into one  
block spin having the length $2 s.$ The fraction of such spins is 
$p\sim \rho R_0^2=(T_0/T)^{2/3}.$ 
The rest of the spins 
are separated by distances  exceeding $R_0$ and can be regarded as 
an ideal gas of spins of length $s.$ 
Then the mixture of the ideal gas of pairs and 
the ideal gas of single spins has the Curie susceptibility
$\chi=  c_0 (1-p)s^2/T + c_0 p (2 s)^2 /2 T,$ which deviates from the ideal 
gas susceptibility, $c_0 s^2 /T,$ by a non-analytic correction 
$\delta \chi \propto p/T\propto (T)^{-5/3}.$ 

{\it  A quantitative theory for the Dirac-RKKY systems in paramagnetic phase}  requires
a resummation of the short-distance singularities appearing in the 
disorder average of the observables.  This is achieved by combining the replica method with 
the virial expansion of the free energy in the temperature-dependent gas parameter $p.$  Thermodynamic 
properties of the system are encoded in the potential 
\begin{equation}
F(h,T,N;q)=-T \ln  \overline{[Z (h,T,N)]^q},
\label{F}
\end{equation} 
where $Z(h,T,N)$ is the  partition function of a given realization of the 
system of $N$ impurities at tempearture $T$ and  
in the presence of the magnetic field $h.$  The integer $q$ is the number of identical 
replicas of the disordered system. The overline stands for the averaging over all quenched variables,
\begin{equation}
\overline{ f }\equiv \frac{1}{\Xi^N \mathcal A^N} 
\sum_{\xi_1\dots \xi_N} \int \prod_{i=1}^N [d\mathbf r_i]f(\mathbf r_1, \dots, \mathbf r_N, \xi_1, \dots, \xi_N),
\notag
\end{equation}
where $\mathcal A$ is the area of the sample, and 
$\Xi=\sum_{\xi} 1$ is the number of distinct values of the variable $\xi.$ 

The magnetic susceptibility \eqref{chidef} can be written as  
\begin{equation}
\chi(T)= \frac{c_0}{N}\lim_{h\to 0} \lim_{q\to0} \frac{\partial^2}{\partial h^2}
\frac{\partial }{\partial q}F(h,T,N;q) ,
\label{chiav}
\end{equation}
which requires analytic continuation of the potential $F$ to the positive real axis of 
$q.$  In order to obtain the virial expansion for the 
susceptibility \eqref{chiav} it is convenient to consider  
the grand canonical ensemble and introduce the thermodynamic potential 
\begin{equation} 
\Omega=-T \ln\left[\sum_{N=0}^{\infty} \frac{e^{\frac {\mu N}{T} }}{N!}
\overline {Z^q (h,T,N)} \right] =
-T   \sum_{n=1}^\infty \frac{ V_n}{ n!}e^{\frac {\mu n}{T} },
\notag
\end{equation}
which is related to the potential \eqref{F} by the Legendre transformation,
\begin{equation}
F=\Omega+\mu N,  \qquad N=-\frac{\partial \Omega}{ \partial \mu}.
\end{equation}
The chemical potential $\mu$ is a $q$-dependent auxiliary variable, which  does 
not have any straightforward physical meaning. The coefficients  $V_n$ 
in $\Omega$ are called the virial coefficients. The first three
of those are:
\begin{eqnarray}
\notag
V_1&=&\overline{Z_1^q} \\
\notag
V_2&=&
\overline{Z_{12}^q- Z_1^q Z_2^q}\\
\notag
V_3&=&\overline{\hat S [Z_{123}^q - 3 Z_{12}^q Z_3^q +2 Z_1^q Z_2^q Z_3^q]} 
\notag\\
\notag
\\
&&Z_{i_1\ldots i_n}=\mathrm{ Tr}e^{-\beta H_{i_1\ldots i_n}}
\end{eqnarray}
where $Z$ 
is a function of $n$ quenched variables $\{\mathbf r_{i_1}, \dots , \mathbf r_{i_n} \}$ 
and $\{\xi_{i_1}, \dots , \xi_{i_n} \}$   describing  localized spins with 
indices $i_1, \dots, i_n\in\{1, \dots, N\},$ and $H_{i_1, \dots, i_n}$ is the Hamiltonian 
\eqref{DHam} constrained to this $n$-subset. The trace is taken over all spin variables
of the $n$-subset, and the symbol $\hat S$ represents a complete 
symmetrization of the expression over the particle indices. Note that $\hat S$ does 
not appear in $V_1$ and $V_2$ because these expressions are already symmetric.

After substituting the virial expansion of $\Omega$  
into \eqref{chiav} one arrives at Eq.~\eqref{answer},
with 
\begin{multline}
C_n=\left[\frac{3}{s(s+1)} \right]^{1+\frac{2}{3}(n-1)}\frac{1}{\Xi^n n!} \\ \times \sum_{\xi_1, \dots \xi_n} 
\int d\mathbf x_2 \dots d\mathbf x_n \hat S \left. \frac{d^n}{dz^n} Q(z)\right \vert_{z=0},
\label{Cdef}
\end{multline}
where the integration is over $n-1$ dimensionless variables 
$(\mathbf x_2, \dots \mathbf x_n)=(\mathbf r_2, \dots, \mathbf r_n)/R_0 .$
The generating function $Q(z)$ in Eq.~\eqref{Cdef} is defined by the infinite series 
\begin{align}
 Q(z)= e^{-z} \left(z M_1+ \frac{z^2}{2!} M_{12} + \frac{z^3}{3!} M_{123}  +\ldots\right),
\notag
 \\
M_{1\dots n} =\frac{\mathrm{Tr} (s_1^z+\dots+s_n^z)^2 e^{-\sum_{i<j}
\frac{G_{ij}}{\vert \mathbf x_i-\mathbf x_j\vert^3 } } }{
\mathrm{Tr}\, e^{-\sum_{i<j}
\frac{G_{ij}}{\vert \mathbf x_i-\mathbf x_j\vert^3 } } }.
\label{Qdef}
\end{align}
Note that there is no integration over $\mathbf x_1$ in Eq.~\eqref{Cdef}, and the 
answer does not depend on the choice of $\mathbf x_1$ by translational invariance of the problem.
Equations \eqref{answer}, \eqref{Cdef}, and \eqref{Qdef}
constitute the main result of this work, which can be applied to any random 
Dirac-RKKY model.

{\it High-temperature expansion and the critical point of the Ising model with random Dirac-RKKY exchange.}
As an illustration, we consider the paramagnetic susceptibility of 
the ferromagnetic Ising model 
described by the pairwise interaction function $
G_{ij}=-s^z_i s^z_j,
$
where $s_i^z= \pm 1/2$ is the Ising spin of the $i$-th atom.
This model describes, for example,  an ensemble of easy-axis magnetic impurities 
on the sufrace of a topological insulator \cite{Abanin}; it also describes structural 
phase transitions in graphene \cite{Structural}.  The only random 
quenched variables in this model are the positions of impurities in the sample. The first two virial 
coefficients found from Eqs.~\eqref{Cdef}, \eqref{Qdef} are 
\begin{eqnarray}
C_1&=&1 \notag \\ \notag
C_2 &=& \frac{ 1}{2} \int_0^\infty  \left[\frac{4}{1+e^{-\frac{2}{x^3}}}-2\right] 2\pi  x  d x=9.01845\dots
\end{eqnarray}
Similarly, all higher-order $C_n$ are expressed as integrals of elementary functions of increasing 
complexity. The numerical values of the first twelve coefficients $C_n$ 
are given in Table~\ref{I}.


\begin{table}
\begin{center}
  \begin{tabular}{| l | l || l | l || l | l |  }
    \hline
$C_1$                             & 1         &  $C_5$  & $3.128(2)\times10^{3}$      & $C_9$  & $5.612(4)\times 10^{6}$
\\ \hline
$C_2$                             & 9.0184534\dots  & $C_6$  & $ 2.103(1)\times 10^{4}$  &  $C_{10} $  &  $3.328(3)\times 10^{7}$
\\
    \hline
$ C_3$   & $6.578(2)\times 10^{1}$ & $C_7$  & $ 1.3937(5)\times 10^{5}$  &  $C_{11}$  & $ 1.857(3) \times 10^{8}$ 
\\ \hline
$C_4$    & $4.582(1)\times 10^{2} $ &  $C_8$  & $ 9.007(4)\times 10^{5}$ & $C_{12}$ &  $ 1.048(7)\times 10^{9}$
\\ \hline
\end{tabular}
\end{center}
\caption{Numerical values of the coefficients in the high-temperature expansion 
\eqref{answer} for the susceptibility of the ferromagnetic Ising model.\label{I}}  
\end{table}

At low enough temperature the Ising model with a ferromagnetic exchange undergoes a phase 
transition into a ferromagnetically ordered state. In order to extract the properties of the critical 
point from the high-temperature data in Table~\ref{I} 
we use the D-log Pad\'e technique \cite{Baker1975} to investigate the function 
$f(z)=\sum_{n\ge 1} C_n z^{n-1},$ where $C_n$ are the same 
coefficients as appear in Eq.~\eqref{answer}, and $z=(T_0/T)^{2/3}.$  More precisely, we construct the first few 
$[L/L+1]$ Pad\'e approximants, $0<L\le 5,$  to the function $g(z)= f'(z)/f(z)$ and analyse the 
structure of the singularity of the approximants in the vicinity of the hypothetical critical point.
For the $L=5$ approximant, which requires the knowledge of the first twelve 
coefficients $C_n,$ we solve the differential equation $ d\ln[ f(z)]/ dz = [L/L+1]_g(z)$ with the initial 
condition $f(0)=0$ and compare the result with the susceptibility obtained from Monte Carlo
simulations on an ensemble of $10^5$ impurities averaged over 100 configurations of quenched 
disorder, see Fig.~\ref{Fig1}. The simulations are performed with the classical worm algorithm \cite{worm} using the ALPS
 libraries \cite{alps}. 
Since the numerical values of $C_n$  have been calculated with finite precision, 
the position of the singularity of each approximant is found with some uncertainty (see the inset 
of Fig.~\eqref{Fig1}). For the $L=5$ approximant we find $T_c/T_0=14.5(5),$ which is 
close to the value found from numerical simulations in \cite{Structural,Abanin,Rappoport}.
We also extract the value of the susceptibility critical exponent $\gamma$ by calculating 
the residue of the $L=5$ Pad\'e approximant at the critical point and find that $\gamma=1.4(2).$

We conclude this report with the discussion of several other examples of the Dirac-RKKY 
systems identified in earlier literature. A detailed analysis of criticality in 
such systems is beyond the scope of this letter, however there is enough 
interesting information already contained in the leading coefficients $C_1$ and $C_2$ 
of the expansion \eqref{answer}. Not only 
do these coefficient describe an experimentally measurable deviation 
of $\chi(T)$ from the Curie law, but also they can be used to 
extract a generalized Curie temperature
$\Theta = {\rm sign}( C_1)\times (|C_2|/C_1)^{3/2} T_0, $ which gives an 
estimate of $T_c$ if the susceptibility is calculated for the 
observable corresponding to the order parameter.  
For spin systems having competing orders one can compare the 
generalized Curie temperatures extracted from the susceptibilities 
for the suspected order parameters for which ordering is 
more likely to occur. Such invormation is given in Table~\ref{II}. The first two rows 
describe two limiting cases of a more general model derived in \cite{Abanin}
for the RKKY exchange in chiral metals. The authors of \cite{Abanin} observe
that a strongly anisotropic in-plane exchange tends to destroy 
the magnetic order enforced by the out-of-plane ferromagnetic coupling.  They 
also conjecture that as a function of the anisotropy parameter the system 
undergoes a quantum phase transition into a spin-glass state. We do not find 
any signature of such a transition in the generalized Curie temperature for both 
in- and out-of-plane susceptibility.

The third row of Table~\ref{II} describes a Dirac-RKKY model of  
magnetic adatoms at the centers of carbon hexagons in graphene \cite{Sherkunov}. The 
quenched parameter $\xi$ labels the three distinct positions of the atom in the 
$\sqrt 3\times \sqrt 3$ superlattice formed by the commensurate Friedel oscillations 
of the RKKY interaction. In this case ferromagnetic ordering is replaced by a
staggered state with the order parameter 
$$\Phi = \sum_i [s^x_i \cos( 2\pi \xi_i/3)+ s^y_i \sin (2\pi \xi_i/3) ].$$ 

Together with the detailed analysis of the Ising system these examples demonstrate 
how the proposed generic high-temperature expansion, Eqs.\eqref{answer}, \eqref{Cdef} and \eqref{Qdef}
can be used to describe the magnetic properties of disordered Dirac-RKKY systems.

 We thank I. Aleiner, B. Altshuler and M. Mezard for helpful discussions, 
and ERC, EPSRC and Royal Society for financial support.

\bibliographystyle{apsrev}

\begin{thebibliography}{100}
\bibitem{footnote1} There may be more than one equilibrium position of an adatom in the 
unit cell of the host. For example, an adatom binding covalently to a carbon atom in graphene 
will have to chose between the A and B sublattices.
\end{thebibliography}

\end{document}